\documentclass[english,pra,aps,amsmath,amssymb,showpacs,preprint]{revtex4}
\usepackage[T1]{fontenc}
\usepackage[latin9]{inputenc}
\usepackage{amsmath}
\usepackage{amssymb}
\usepackage{esint}

\makeatletter
\@ifundefined{textcolor}{}
{%
 \definecolor{BLACK}{gray}{0}
 \definecolor{WHITE}{gray}{1}
 \definecolor{RED}{rgb}{1,0,0}
 \definecolor{GREEN}{rgb}{0,1,0}
 \definecolor{BLUE}{rgb}{0,0,1}
 \definecolor{CYAN}{cmyk}{1,0,0,0}
 \definecolor{MAGENTA}{cmyk}{0,1,0,0}
 \definecolor{YELLOW}{cmyk}{0,0,1,0}
}

\usepackage{subfigure}\usepackage{hyperref}\usepackage{float}\hypersetup{
unicode=false,
pdftoolbar=true,
pdfmenubar=true,
pdffitwindow=false,
pdfstartview={FitH},
pdftitle={Mytitle},
pdfauthor={Author},
pdfsubject={Subject},
pdfcreator={Creator},
pdfproducer={Producer},
pdfkeywords={keywords},
pdfnewwindow=true,
colorlinks=true,
linkcolor=red,
citecolor=blue,
filecolor=magenta,
urlcolor=cyan
}

\makeatletter\newcommand{\Rmnum}[1]{\expandafter\@slowromancap\romannumeral #1@}\makeatother

\makeatother

\usepackage{babel}

\makeatother

\usepackage{babel}

\makeatother

\usepackage{babel}

\makeatother

\usepackage{babel}

\makeatother

\usepackage{babel}

\makeatother

\usepackage{babel}

\makeatother

\usepackage{babel}

\makeatother

\usepackage{babel}

\makeatother

\usepackage{babel}

\makeatother

\usepackage{babel}

\makeatother

\usepackage{babel}
\begin{document}

\title{Berry phases for interacting spins in composite environments }

\author{Da-Bao Yang }

\affiliation{Theoretical Physics Division, Chern Institute of Mathematics, Nankai
University, Tianjin 300071, People's Republic of China}

\author{Fu-Lin Zhang}

\affiliation{Physics Department, School of Science, Tianjin University, Tianjin
300072, People's Republic of China}

\author{Jing-Ling Chen}

\email{chenjl@nankai.edu.cn}

\affiliation{Theoretical Physics Division, Chern Institute of Mathematics, Nankai
University, Tianjin 300071, People's Republic of China}

\affiliation{Centre for Quantum Technologies, National
University of Singapore, 3 Science Drive 2, Singapore 117543}

\date{\today}
\begin{abstract}
Due to the potential application in quantum information process, geometric
phase of interacting system arouse many interests. Some physicists
concentrate on the system in pure classical environment, while others
study the system in pure quantized environment. So a natural question
is asked: how about an interacting system in composite environments
made up of both classical and quantized field. In this letter, we
analyze a quantum system composed of two interacting spins, of which
one is in classical magnetic field and the other is in quantized field.
First, classical magnetic field driven Berry phases for the whole
system and subsystem are studied. The effect of couplings between
particles and photon on these phases are analyzed. In comparison with
the dynamical quantized field, We find that even a static quantized
field in its vacuum state can also have an effect on Berry phase.
Second, quantized field driven Berry phases for the whole system and
subsystem are formulated, including both one and two mode of this
field. The vacuum induced effects are elaborated, moreover compared
with the constant vacuum induced phases in former papers, the counterpart
in this letter varies according to classical magnetic field, couplings
and other parameters. For the two mode quantized field, the rigorous
relationship between the concurrence and Berry phase for subsystem
are built up.
\end{abstract}


\maketitle

\section{Introduction}

\label{sec:introduction}

Berry phase had been discovered by Berry \cite{berry1984quantal}
in the context of adiabatic, unitary, cyclic evolution of time-dependent
quantum system. He demonstrated that besides the usual dynamical phase,
there exists an additional phase relating to the evolution of the
state. Soon A geometrical interpretation of Berry phase was elaborated
by Simon \cite{simon1983holonomy}. Berry's result was extended to
the nonadiabatic and cyclic case by Aharonov and Anandan \cite{aharonov1987phase}.
In their definition, the dynamical phase was identified as a loop
integral over the expectation value of the Hamiltonian. The Aharonov
and Anandan phase (A-A phase) could be obtained by the difference
between the total phase and the dynamical one. Likewise, Anandan \cite{anandon1988nonadiabatic}
generalized the above one to the degenerate case. Depending on the
Pancharatnam's earlier work \cite{pancharatnam1956connection}, Samuel
and Bhandari \cite{samuel1988general} found a more general phase
in the context of non-cyclic and non-unitary evolution of quantum
mechanics. Subsequently, Munkunda and Simon established a quantum
kinematic approach to geometric phases \cite{mukunda1993quantum},
which is the most general theory on geometric phases for pure quantum
states.

Nevertheless, the above definitions of geometric phases can't be applicable
when the initial and final states are orthogonal. Manini and Pistolesi
\cite{manini2000off} first proposed the Abelian off-diagonal geometric
phases to overcome the evident drawback of ordinary Berry phase during
adiabatic evolution. One year later, the above definition was generalized
to nonadiabatic cases by Mukunda et. al.\cite{mukunda2001bargmann}.
Afterwards, Kult \emph{et al}. \cite{kult2007nonabelian} made a step
forward in this direction by extending the concepts to non-Abelian
cases.

The above definitions are all confined to pure states, what about
the geometric phase for mixed state? This problem was first presented
by Uhlmann \cite{uhlmann1986parallel} in mathematical context of
purification. But, his definition depends on the chosen environment.
Later, Sj$\ddot{o}$qvist \emph{et al}. \cite{sjoqvist2000mixed}
redefine the non-degenerate mixed geometric phase in non-cyclic and
unitary evolution under the background of quantum interference, which
is independent of surroundings. Extensions of mixed-state geometric
phases to the degenerate case \cite{singh2003geometric} and the kinematic
approach \cite{tong2004kinematic} had also been achieved. Likewise,
the definition of mixed geometric phase was also extended to off-diagonal
case by Filipp and Sj$\ddot{o}$qvist.

Besides its theoretical significance, Berry phase has many applications
ranging from condensed matter physics \cite{xiao2010Berry} to quantum
information and computation science \cite{jones1999geometric,duan2001geometric,zhu2003unconventional,feng2009geometric}.
Most of the implementation of quantum information by geometric phase
depends on composite system. So this kind of systems is of concern.
Sj$\ddot{o}$vist \cite{sjoqvist2000geometric} studied geometric
phase for a pair of entangled spins in a time-independent uniform
magnetic field, which was generalized by Tong, Kwek and Oh \cite{tong2003geometric}
to a rotating magnetic field. Yi, Wang and Zheng \cite{yi2004berry}
investigated Berry phase two two coupled spin half system, one of
which is driven by a slowly varying magnetic field. Sj$\ddot{o}$qvist
et. al. \cite{sjoqvist2010berry} analyzed Berry phase for ground
state of finite-size Lipkin-Meskov-Glick model including three spin
half particles, which was extended to A-A geometric phase by Yang
et. al. \cite{yang2010aharonov}.

However, all the above researched system were in classical field.
Recently, some scientists began to study spins in quantized field.
Fuentes-Guridi et. al. \cite{fuentes2002vacuum} studied the Berry
phase for spin half particle interacting with a quantized field and
analyzed the vacuum induced effect, which was generalized by Liu et.
al. \cite{liu2010vacuum} to $m$ photons process. Wang, Cui and Yi
\cite{wang2004berry} considered about interacting spins with one
driven by a quantized mode of field, while the counterpart of both
of the two particles driven by the field was analyzed by Liang, Zhang
and Yuan \cite{liang2007berry}. So a natural question is asked: how
about an interacting system in composite environments made up of both
classical and quantized field. In this letter, we analyze a quantum
system composed of two interacting spins, of which one is in classical
magnetic field and the other is in quantized field.

This paper is organised as follows. In the next section, Berry phases
with classical magnetic field driving are discussed. The eigenstates
of the whole system are worked out, moreover they are represented
by triangular function of some introduced parameters, which made them
elegant and easy to understand. Not only the Berry phase for the whole
system but also the Berry phase for subsystem are formulated. Some
special cases at utmost limit are discussed and the concurrence is
used to explain the above phenomena. In Sec. III, the Berry phase
with quantized field driving is calculated, including for both the
whole system and subsystem. The vacuum induced Berry phase are placed
emphasis on. The relationship between the concurrence and Berry phase
for subsystem are built up. Furthermore, some special cases are elaborated.
A conclusion is drawn in the last section.

\section{Berry phase with magnetic field driving}

\label{sec:reviews}

Considering a system consisting of two interacting spin-$1/2$ particles
in the presence of composite fields and supposing that particle 1
interacts with a single quantized mode of an optic field in the rotating
wave approximation and particle 2 is subject to a classical magnetic
field, the Hamiltonian takes the form
\begin{equation}
H=\frac{\omega_{1}}{2}\sigma_{1}^{z}+\nu a^{\dagger}a+\lambda(\sigma_{1}^{+}a+\sigma_{1}^{-}a^{\dagger})+J\sigma_{1}^{z}\sigma_{2}^{z}+\frac{1}{2}\mu\vec{B}\cdot\vec{\sigma}_{2},\label{eq:HamiltonianOneMode}
\end{equation}
where $\omega_{1}$ is the transition frequency between the eigenstates
of particle 1, $\nu$ is the frequency of the field described in terms
of the creation and annihilation operators $a^{\dagger}$ and $a$,
$\lambda$ is the coupling constant between the quantized field and
particle 1, $J$ denotes the coupling constant between the two particles,
$\mu$ represents the gyromagnetic ratio, $\vec{B}=B\vec{n}$ stands
for the magnetic field, $\vec{\sigma_{k}}=(\sigma_{k}^{x},\sigma_{k}^{y},\sigma_{k}^{z})$,
$\sigma_{k}^{+}=(1/2)(\sigma_{k}^{x}+i\sigma_{k}^{y})$ and $\sigma_{k}^{-}=(1/2)(\sigma_{k}^{x}-i\sigma_{k}^{y})$
are Pauli operators, the subscript denotes the particle.

In the invariant space spanned by $\{|e_{1}e_{2}n\rangle,|e_{1}g_{2}n\rangle,|g_{1}e_{2}n+1\rangle,|g_{1}g_{2}n+1\rangle\}$,
the Hamiltonian \eqref{eq:HamiltonianOneMode} can be expressed in
a matrix form,
\[
\left(\begin{array}{cccc}
J+n\nu+\frac{\omega_{1}}{2} & \frac{1}{2}e^{-i\varphi}\omega_{2} & \sqrt{n+1}\lambda & 0\\
\frac{1}{2}e^{i\varphi}\omega_{2} & -J+n\nu+\frac{\omega_{1}}{2} & 0 & \sqrt{n+1}\lambda\\
\sqrt{n+1}\lambda & 0 & -J+(n+1)\nu-\frac{\omega_{1}}{2} & \frac{1}{2}e^{-i\varphi}\omega_{2}\\
0 & \sqrt{n+1}\lambda & \frac{1}{2}e^{i\varphi}\omega_{2} & J+(n+1)\nu-\frac{\omega_{1}}{2}
\end{array}\right),
\]
where $\vec{n}=(\cos\varphi,\sin\varphi,0)$ and $\omega_{2}=\mu B$.
And the four nondegenerate eigenvectors are
\begin{equation}
|\psi_{j}\rangle=\begin{array}{cccc}
(e^{-i\varphi}\cos\frac{\chi_{j}}{2}\sin\frac{\xi_{j}}{2} & \cos\frac{\chi_{j}}{2}\cos\frac{\xi_{j}}{2} & e^{-i\varphi}\sin\frac{\chi_{j}}{2}\sin\frac{\eta_{j}}{2} & \sin\frac{\chi_{j}}{2}\cos\frac{\eta_{j}}{2}\end{array})^{T},\label{eq:InstantaneousEigenvecor}
\end{equation}
where
\[
\begin{array}{ccc}
\cos\frac{\chi_{j}}{2}\sin\frac{\xi_{j}}{2} & = & \frac{1}{\sqrt{N}}(\omega_{1}\{2J[\sqrt{F\mp2A}+(-1)^{j}2(J-\nu)]+(-1)^{j}(\omega_{2}^{2}\mp A)\}\\
 &  & +(A\pm2J\nu)[\mp\sqrt{F-2A}+(-1)^{j}(\pm\nu\mp2J)]+(-1)^{j}(2J\omega_{1}^{2}-\nu\omega_{2}^{2}))\\
\cos\frac{\chi_{j}}{2}\cos\frac{\xi_{j}}{2} & = & \frac{1}{\sqrt{N}}(\omega_{2}\{\omega_{1}[\sqrt{F\mp2A}+(-1)^{1+j}2\nu]-\nu[\sqrt{F\mp2A}+(-1)^{1+j}\nu]\\
 &  & +(-1)^{j}(4\lambda_{n}^{2}+\omega_{1}^{2}\mp A)\})\\
\sin\frac{\chi_{j}}{2}\sin\frac{\eta_{j}}{2} & = & \frac{1}{\sqrt{N}}[(-1)^{j}2\lambda_{n}\left(2J\omega_{1}+\omega_{2}^{2}\mp A-2J\nu\right)]\\
\sin\frac{\chi_{j}}{2}\cos\frac{\eta_{j}}{2} & = & \frac{1}{\sqrt{N}}\{2\omega_{2}\lambda_{n}[\sqrt{F\mp2A}+(-1)^{j}2J]\}
\end{array},
\]
 the normalized coefficients are
\[
\begin{array}{ccc}
N_{j} & = & (\omega_{1}\{2J[\sqrt{F\mp2A}+2(-1)^{j}(J-\nu)]+(-1)^{j}\left(\mp A+\omega_{2}^{2}\right)\}\\
 &  & +(A\pm2J\nu)[\mp\sqrt{F\mp2A}\mp(-1)^{j}(\nu-2J)]+(-1)^{j}\left(2J\omega_{1}^{2}-\nu\omega_{2}^{2}\right)){}^{2}\\
 &  & +4\omega_{2}^{2}\lambda_{n}^{2}[\sqrt{F\mp2A}+2(-1)^{j}J]^{2}+\omega_{2}^{2}[-(-1)^{j}\nu\sqrt{F\mp2A}+(-1)^{j}\omega_{1}\sqrt{F\mp2A}\\
 &  & \mp A+\left(\nu-\omega_{1}\right){}^{2}+4\lambda_{n}^{2}]^{2}+4\lambda_{n}^{2}\{[A\pm2J\left(\nu-\omega_{1}\right)]\mp\omega_{2}^{2}\}{}^{2}
\end{array},
\]
$j=1,2,3,4$ (when $j=1,2$, the above sign is sensible; other cases,
the bellow sign is sensible), $A=\sqrt{4J^{2}\left(\nu-\omega_{1}\right){}^{2}+\omega_{2}^{2}[\nu^{2}+\omega_{1}\left(\omega_{1}-2\nu\right)+4\lambda^{2}(n+1)]}$,
$F=4J^{2}+\nu^{2}+\omega_{1}\left(\omega_{1}-2\nu\right)+4\lambda^{2}(n+1)+\omega_{2}^{2}$,
and $\lambda_{n}=\lambda\sqrt{n+1}$.

When $\varphi$ is slowly changed from $0$ to $2\pi$, the system
undergoes an adiabatic and cyclic evolution. Then, the corresponding
Berry phase can be calculated as the following formula,
\begin{equation}
\gamma_{j}=i\intop_{0}^{2\pi}d\varphi\langle\psi_{j}|\frac{d}{d\varphi}|\psi_{j}\rangle.\label{eq:BerryPhase}
\end{equation}
Substituting the instantaneous eigenvector \eqref{eq:InstantaneousEigenvecor}
into the above Eq. \eqref{eq:BerryPhase}, one can obtain Berry phase,
\begin{equation}
\frac{1}{2}[\sin^{2}(\frac{\chi_{j}}{2})\Omega(\eta_{j})+\cos^{2}(\frac{\chi_{j}}{2})\Omega(\xi_{j})],\label{eq:BerryPhaseForMagneticDriving}
\end{equation}
where
\begin{equation}
\Omega(x)=2\pi(1-\cos x).\label{eq:SolidAngle}
\end{equation}
The above Eq. \eqref{eq:SolidAngle} hints that $\Omega(\eta_{j})$
and $\Omega(\xi_{j})$ can be regarded as the solid angle of a sphere
of fix latitudes $\eta_{j}$ and $\xi_{j}$ respectively. So the geometric
phase \eqref{eq:BerryPhaseForMagneticDriving} is a linear combination
of solid angles. And $\sin\frac{\chi_{j}}{2}$ and $\cos\frac{\chi_{j}}{2}$
play the roles of weights. References \cite{fuentes2002vacuum,wang2004berry,liang2007berry,liu2010vacuum}
had investigated that when $n=0$, the time-dependent quantized field
induced a corresponding Berry phase. However, in this paper, we get
another conclusion that even if the quantized field is static, when
$n=0$, the vacuum quantum field had an effect on Berry phase by the
coupling constant $\lambda$ or the frequency $\nu$ of the quantized
field. Moreover, if either $\lambda$ or $\nu$ is also zero, the
corresponding Berry phase is also affected by $\nu$ or $\lambda$.
The above discussion focus on the affection of quantized filed. Now,
let's turn our interests into the impaction of classical magnetic
field. When $\vec{B}=0$, i.e. , $\omega_{2}=0$, the Berry phase
\eqref{eq:BerryPhaseForMagneticDriving} disappears as we expect.
Because when $\vec{B}=0$, the instantaneous eigenstates become stationary.
Therefore, no geometric phase can be generated. Next, let's concentrate
on some special cases. Under the condition that the coupling constant
$\lambda\rightarrow\infty$ or $n\rightarrow\infty$, the geometric
phase
\begin{equation}
\gamma_{j}=\pi.\label{eq:BerryPhaseParametersGoesInfinity}
\end{equation}
In order to explain the above result, the concurrence of the instantaneous
eigenstates is calculated as follows
\begin{equation}
c_{j}=\sin(\chi_{j})\sin\frac{\xi_{j}-\eta_{j}}{2}.\label{eq:concurrence}
\end{equation}
It can be verified that under the condition that $\lambda\rightarrow\infty$
or $n\rightarrow\infty$, the related concurrence becomes zero. Hence,
there is no relationship between particle 1 and particle 2. And the
state of particle 1 is stationary. So the whole geometric phase is
only generated by particle 2 precessing in the classical magnetic
field. On the assumption of $J\rightarrow\infty$, the geometric phase

\[
\gamma_{j}=0.
\]
 It may be explained as that when $J\rightarrow\infty$, the interaction
between spins in Hamiltonian \eqref{eq:HamiltonianOneMode} becomes
dominate part, hence the effect of classical magnetic field is negligible.
Therefore, the Berry phase generated by the precession of the field
is null. When $J=0$, the concurrence vanishes. As the same circumstance
above that $\lambda\rightarrow\infty$ or $n\rightarrow\infty$, the
corresponding Berry phase is $\pi$.

Next, let us research mixed geometric phases for subsystems, which
is proposed in \cite{sjoqvist2000geometric}
\begin{equation}
\Gamma=\arg\sum_{l}p_{l}e^{i\beta_{u}},\label{eq:MixedStateGeometricPhase}
\end{equation}
where $p_{l}$ and $\beta_{l}$ are the reduced density matrix's eigenvalues
and Berry phases generated by the eigenvectors respectively. The reduced
density matrix of particle 2 is
\[
\left(\begin{array}{cc}
\cos^{2}(\frac{\chi_{j}}{2})\sin^{2}\left(\frac{\xi_{j}}{2}\right)+\sin^{2}\left(\frac{\eta_{j}}{2}\right)\sin^{2}(\frac{\chi_{j}}{2}) & \frac{1}{2}e^{-i\varphi}[\sin(\xi_{j})\cos^{2}(\frac{\chi_{j}}{2})+\sin(\eta_{j})\sin^{2}(\frac{\chi_{j}}{2})]\\
\frac{1}{2}e^{i\varphi}[\sin(\xi_{j})\cos^{2}(\frac{\chi_{j}}{2})+\sin(\eta_{j})\sin^{2}(\frac{\chi_{j}}{2})] & \cos^{2}\left(\frac{\xi_{j}}{2}\right)\cos^{2}(\frac{\chi_{j}}{2})+\cos^{2}\left(\frac{\eta_{j}}{2}\right)\sin^{2}(\frac{\chi_{j}}{2})
\end{array}\right).
\]
 Hence, by use of Eq. \eqref{eq:MixedStateGeometricPhase}, we obtain
the corresponding mixed state geometric phase for particle 2, which
takes the form
\[
\Gamma_{j}=-\tan^{-1}\left(\frac{1}{2}\sqrt{2\sin^{2}(\chi_{j})\cos(\eta_{j}-\xi_{j})+\cos(2\chi_{j})+3}\tan\frac{2\pi[\cos(\eta_{j})\sin^{2}(\frac{\chi_{j}}{2})+\cos(\xi_{j})\cos^{2}(\frac{\chi_{j}}{2})]}{\sqrt{2\sin^{2}(\chi_{j})\cos(\eta_{j}-\xi_{j})+\cos(2\chi_{j})+3}}\right).
\]
 As the geometric phase for the whole system, when the coupling constant
$J=0$, the geometric phase for particle 2 reduces to be
\[
\Gamma_{j}=\pi
\]
 as well.

\section{Berry phase with quantized field driving}

\label{sec:Nonadiabatic}

In this section the phase shift operation $U(\phi)=\exp(-i\phi a^{\dagger}a)$
is introduced to apply adiabatically to the Hamiltonian of the system
\eqref{eq:HamiltonianOneMode}. When $\phi$ is slowly changed from
$0$ to $2\pi$, the corresponding Berry phase can be obtained as
follows:
\begin{equation}
\gamma_{j}^{q}=i\intop_{0}^{2\pi}d\phi\langle\psi_{j}|U^{\dagger}(\phi)\frac{d}{d\phi}U(\phi)|\psi_{j}\rangle.\label{eq:GeometricPhaseForQuantumDriving}
\end{equation}
Substituting the instantaneous state \eqref{eq:InstantaneousEigenvecor}
into the above expression \eqref{eq:GeometricPhaseForQuantumDriving},
we obtain
\begin{equation}
\gamma_{j}^{q}=\pi(1-\cos\chi_{j})+2\pi n.\label{eq:ConcretGeometricPhaseForQuantumDriving}
\end{equation}
The static classical magnetic field $\vec{B}$, the interaction between
the two particles and quantum field all have impacts on the geometric
phase of the system. Even when $n=0$, the vacuum quantum field still
have an effect on $\gamma_{j}^{q}$ through $\cos\chi_{j}$. However,
when the classical magnetic field $\vec{B}=0$, it gives no effect
on the geometric phase. This may be the prominent feature of effect
on geometric phase between the classical field and quantum field.

Moreover, in order to disclose the vacuum induced effect explicitly,
the second mode of the field will introduced in this system \cite{fuentes2002vacuum,wang2004berry},
whose creation and annihilation operators are labeled by $b^{\dagger}$
and $b$ respectively. The Hamiltonian of the whole system takes the
form
\begin{equation}
H_{0}^{2q}=\frac{\omega_{1}}{2}\sigma_{1}^{z}+\nu a^{\dagger}a+\nu b^{\dagger}b+\lambda(\sigma_{1}^{+}a+\sigma_{1}^{-}a^{\dagger})+J\sigma_{1}^{z}\sigma_{2}^{z}+\frac{1}{2}\mu\vec{B}\cdot\vec{\sigma}_{2},\label{eq:HamiltonianTwoMode}
\end{equation}
which implies that the second mode of light has no interaction with
the two spin $1/2$ system or the first mode of light at the initial
time. Hence, the eigenstates of the above Hamiltonian \eqref{eq:HamiltonianTwoMode}
read
\[
|\psi_{j}^{2q}\rangle=|\psi_{j}\rangle\otimes|n^{\prime}\rangle,
\]
where $|\psi_{j}\rangle$ is the eigenstates of the Hamiltonian \eqref{eq:HamiltonianOneMode}
in the previous section. With the help of $b^{\dagger}$ and $b$
of the second mode field, the following transformed Hamiltonian is
considered about
\begin{equation}
U(\theta,\phi)H_{0}^{2q}U^{\dagger}(\theta,\phi),\label{eq:HamiltonianTwoModeTransformed}
\end{equation}
where $U(\theta,\phi)=\exp(-\phi J_{z})\exp(-\theta J_{y})$ whose
generators are $J_{z}=\frac{1}{2}(a^{\dagger}a-b^{\dagger}b)$ and
$J_{y}=\frac{1}{2i}(a^{\dagger}b-ab^{\dagger})$. Therefore, its eigenstates
are
\begin{equation}
U(\theta,\phi)|\psi_{j}^{2q}\rangle.\label{eq:EigenStatesTwoModeTransformed}
\end{equation}
If the parameters $\theta$ and $\phi$ are slowly changed so that
the above eigenstates undertake adiabatic progress, there exist Berry
phases
\begin{equation}
\gamma_{j}^{2q}=\oint\langle\psi_{j}^{2q}|U^{\dagger}(\theta,\phi)dU(\theta,\phi)|\psi_{j}^{2q}\rangle,\label{eq:BerryPhaseTwoModeTransformed}
\end{equation}
where $d$ denote exterior derivative. Substituting Eq. \eqref{eq:EigenStatesTwoModeTransformed}
into Eq. \eqref{eq:BerryPhaseTwoModeTransformed}, one can obtain
the corresponding Berry phases, which take the forms
\begin{equation}
\gamma_{j}^{2q}=-\frac{1}{2}\Omega[(n-n^{\prime})+\sin^{2}\frac{\chi_{j}}{2}],\label{eq:BerryPhaseTwoModeExplicitForm}
\end{equation}
where $\Omega=\intop\sin\theta d\theta\wedge d\phi$ is the solid
angle subtended by the closed loop in the Poincar$\acute{e}$ 's sphere.
From above equation, we can draw a conclusion that even though the
field is in vacuum state, there is also an induced Berry phase, which
is
\begin{equation}
\left(\gamma_{j}^{2q}\right)^{zero}=-\frac{1}{2}\Omega\left(\sin^{2}\frac{\chi_{j}}{2}\right)_{n=0}.\label{eq:VacuumInducedBerryPhaseTwoModeTransformed}
\end{equation}
Compared with the result obtained by Ref. \cite{fuentes2002vacuum}
that the vacuum induced Berry phase is a definite value, our outcomes
\eqref{eq:VacuumInducedBerryPhaseTwoModeTransformed} varies according
to $\sin^{2}\frac{\chi_{j}}{2}$, whose explicit form is
\[
\sin^{2}\frac{\chi_{j}}{2}=\frac{4\lambda_{n}^{2}}{N_{j}}\{\left(2J\omega_{1}+\omega_{2}^{2}\mp A-2J\nu\right)^{2}+\omega_{2}^{2}[\sqrt{F\mp2A}+(-1)^{j}2J]^{2}\}.
\]
We can see that even when $n=n^{\prime}=0$, the quantum filed $\nu$,
the classical field $\textbf{B}$ and the coupling effect $\lambda$
can induce Berry phases. From the first mode and the second mode field,
a comparison is shown that because the second mode field has no interaction
with the particles, when $n^{\prime}=0$, it has no effect on Berry
phase.

The above paragraphs discuss about Berry phase for the whole system.
In the present section, let us concern on the Berry phase of subsystem.
Using the same definition above \eqref{eq:MixedStateGeometricPhase},
the Berry phase for subsystem composed of particle $1$, the first
mode quantum field and the second mode one takes the form
\begin{equation}
\begin{array}{ccc}
\Gamma_{j}^{2q} & = & -\frac{1}{2}\Omega\left[n-n^{\prime}+\frac{1}{2}\right]+\\
 &  & \tan^{-1}\left\{ \sqrt{\sin^{2}(\chi_{j})\cos^{2}\left(\frac{\xi_{j}-\eta_{j}}{2}\right)+\cos^{2}(\chi_{j})}\tan\mathbf{\left[\frac{\Omega\cos(\chi_{j})}{4\sqrt{\sin^{2}(\chi_{j})\cos^{2}\left(\frac{\xi_{j}-\eta_{j}}{2}\right)+\cos^{2}(\chi_{j})}}\right]}\right\} .
\end{array}\label{eq:MixedStateGeometricPhaseTwoMode}
\end{equation}
By substituting Eq. \eqref{eq:concurrence} into above Eq. \eqref{eq:MixedStateGeometricPhaseTwoMode},
we can obtain the relationship between Berry phase of subsystem and
concurrence, which is
\begin{equation}
\Gamma_{j}^{2q}=-\frac{1}{2}\Omega\left[n-n^{\prime}+\frac{1}{2}\right]+\tan^{-1}\left\{ \sqrt{1-c_{j}^{2}}\tan\mathbf{\left[\frac{\Omega\left(1-2\sin^{2}\frac{\chi_{j}}{2}\right)}{4\sqrt{1-c_{j}^{2}}}\right]}\right\} .\label{eq:MixedStateGeometricPhaseTwoModeConcurrence}
\end{equation}
When the corresponding concurrence $c_{j}$ is zero ( For example,
when $J=0$, $c_{j}=0$. ), the Berry phase takes the form
\[
\left(\Gamma_{j}^{2q}\right)_{c_{j=0}}=-\frac{1}{2}\Omega\left[n-n^{\prime}+\sin^{2}\frac{\chi_{j}}{2}\right],
\]
which is very similar to the form \eqref{eq:BerryPhaseTwoModeExplicitForm}.

\section{Conclusions and Acknowledgements }

\label{sec:discussion}

This letter concentrates on a quantum system composed of two interacting
spins, of which one is in classical magnetic field and the other is
in quantized field. First, classical magnetic field driven Berry phases
for the whole system and subsystem are studied. The effect of couplings
between particles and photon on these phases are analyzed. In comparison
with the dynamical quantized field, We find that even a static quantized
field in its vacuum state can also have an effect on Berry phase.
Second, quantized field driven Berry phases for the whole system and
subsystem are formulated, including both one and two mode of this
field. The vacuum induced effects are elaborated, moreover compared
with the constant vacuum induced phases in former papers, the counterpart
in this letter varies according to classical magnetic field, couplings
and other parameters. For the two mode quantized field, the rigorous
relationship between the concurrence and Berry phase for subsystem
are built up.

F.L.Z. is supported by NSF of China under Grant No. 11105097. And
J.L.C. is supported by National Basic Research Program (973 Program)
of China under Grant No. 2012CB921900 and NSF of China (Grant Nos.
10975075 and 11175089). This work is also partly supported by National
Research Foundation and Ministry of Education , Singapore (Grant No.
WBS: R-710-000-008-271).

\end{document}